\documentclass[twocolumn, pra]{revtex4-1}
\usepackage{amssymb}
\usepackage{amsmath}
\usepackage{epsfig}
\usepackage{graphics, graphicx}
\usepackage{bbold}
\usepackage{psfrag}
\usepackage{mathcomp}
\usepackage{subfigure}
\usepackage{verbatim}
\usepackage{color}
\usepackage[colorlinks,citecolor=blue]{hyperref}

\begin{document}

\title{Landau Level to Efimov-like Bound State Crossover in Two-dimensional Dirac Semimetals}
\author{Mingyuan Sun}
\email{msun@connect.ust.hk}
\affiliation{Institute for Advanced Study, Tsinghua University, Beijing, 100084, China}
\date{\today}

\begin{abstract}
 In two-dimensional Dirac semimetals, massless Dirac fermions can form a series of Efimov-like quasibound states in an attractive Coulomb potential. In an applied magnetic field, these quasibound states can become bound states around the Dirac point, due to the energy gap between two successive Landau levels. Through the calculation of the energy spectrum directly, we show that there exists a crossover from the Landau level to the bound state as the magnetic field increases. Because of the decreasing magnetic length, the deep quasibound states are pushed up to the Dirac point relatively and transformed into bound states gradually. Furthermore, the magnetic positions of the emerging bound states in the crossovers obey a geometric scaling law, which can be regarded as an analogy of the radial law in the Efimov effect. We extend our analysis to massless three-component fermions, which also demonstrate this phenomenon. Our results show the universality of fermions with linear dispersion in two dimension and pave a way for the future experimental exploration.
\end{abstract}
\maketitle

\section{Introduction}

Two-dimensional (2D) Dirac semimetal is a good platform to study new physical phenomena \cite{Dirac1,Dirac2,Dirac3}. One famous example is the observation of quantum hall effect in graphene \cite{QHE}. Due to the generally large "fine structure constant" (for example, $\alpha \approx 2$ in graphene, compared with $\alpha=1/137$ in an atomic nucleus), it is also much easier to study the physics of atomic collapse in a strong Colomb potential, which is a fundamental quantum relativistic phenomenon \cite{AC1,AC2,AC3,AC4,AC5,AC6,AC7,AC8,AC9,AC10,AC11,AC12}. Because it is gapless, no bound state can exist in an attractive Colomb potential. However, a sequence of quasibound states can form when the Coulomb potential exceeds a critical value \cite{AC4,AC6,AC12,DiracEfimov1,DiracEfimov2,DiracEfimov3,DiracEfimov4}. These quasibound states follow a geometric scaling law with a universal scaling factor, i.e., exhibiting a discrete scaling symmetry, similar to the Efimov effect \cite{Efimov1,Efimov2}. In this sense, we call them Efimov-like quasibound states in this paper.  

\begin{figure}[b] 
    \includegraphics[width=5.50cm,height=6.0cm]{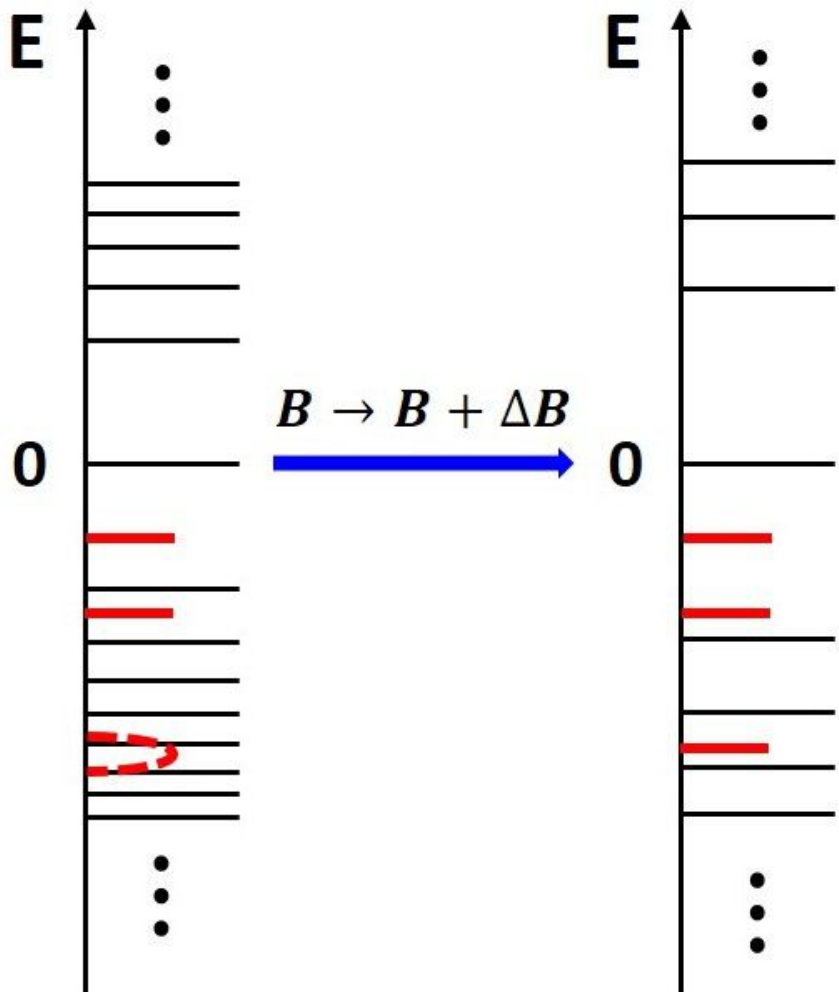}
    \caption{(Color Online). Schematics of the interplay between Landau levels and bound (quasibound) states in 2D Dirac semimetals with a magnetic field and a Coulomb potential. Black lines represent Landau levels which follow the law $E_n \propto \sqrt{n B}$ for free Dirac fermions. Red solid lines denote bound states formed in between Landau levels, while red dashed lines denote quasibound states, whose widths are bigger than the nearby gap between Landau levels. When the magnetic field increases as well as the gap, Landau levels can meet (or cross) bound states (shown on the right side). Hence, there can exist a crossover from the Landau level to the bound state. Moreover, quasibound states can transform into bound states when their widths are smaller than the gap.}
     \label{schematics}
\end{figure}

When a magnetic field is applied, new phenomena occur due to the interplay between Landau levels and Efimov-like quasibound states \cite{AC11,AC12,Sun}. A new length scale, i.e., magnetic length $l_B=\sqrt{\hbar/(eB)}$, is implemented into the system and it breaks quasibound states' original discrete scaling symmetry. However, in the space-magnetic length domain, the discrete scaling symmetry recovers when we transform $r\rightarrow \eta r$ and $l_B\rightarrow \eta l_B$ ($\eta$ is a constant) simultaneously \cite{Sun}. Owing to the gap between two neighbor Landau levels, the quasibound state can become a bound state if the gap is bigger than the width of the quasibound state. Since the gap between Landau levels turns smaller as it is away from the Dirac point, a general scenario should be that bound states form around the Dirac point while quasibound states remain at fairly below, as displayed in Fig.~\ref{schematics}. When the magnetic field increases, more and more deep quasibound states are pushed up relatively to Landau levels and become bound states. A remaining question is that how these bound states interact with Landau levels and modify the energy spectrum.

In this paper, we investigate this problem by calculating the energy spectrum directly. We find that a crossover from a Landau level to a bound state occurs when the Landau level meets the bound state, driven by the varying magnetic field. The magnetic positions for different crossovers obey a geometric scaling law, which is similar to the radial law of the Efimov effect. Therefore, we call them Efimov-like bound states as we do for the quasibound state. We want to emphasize that these bound states only exhibit a discrete scaling symmetry in the radial direction of the space-magnetic length domain. This analysis is applied to massless three-component fermions, where the same physical scenario occurs. Our results indicate that this is a universal phenomenon in 2D fermions with linear dispersion.

\begin{figure}[b] 
    \includegraphics[width=8.0cm,height=13.0cm]{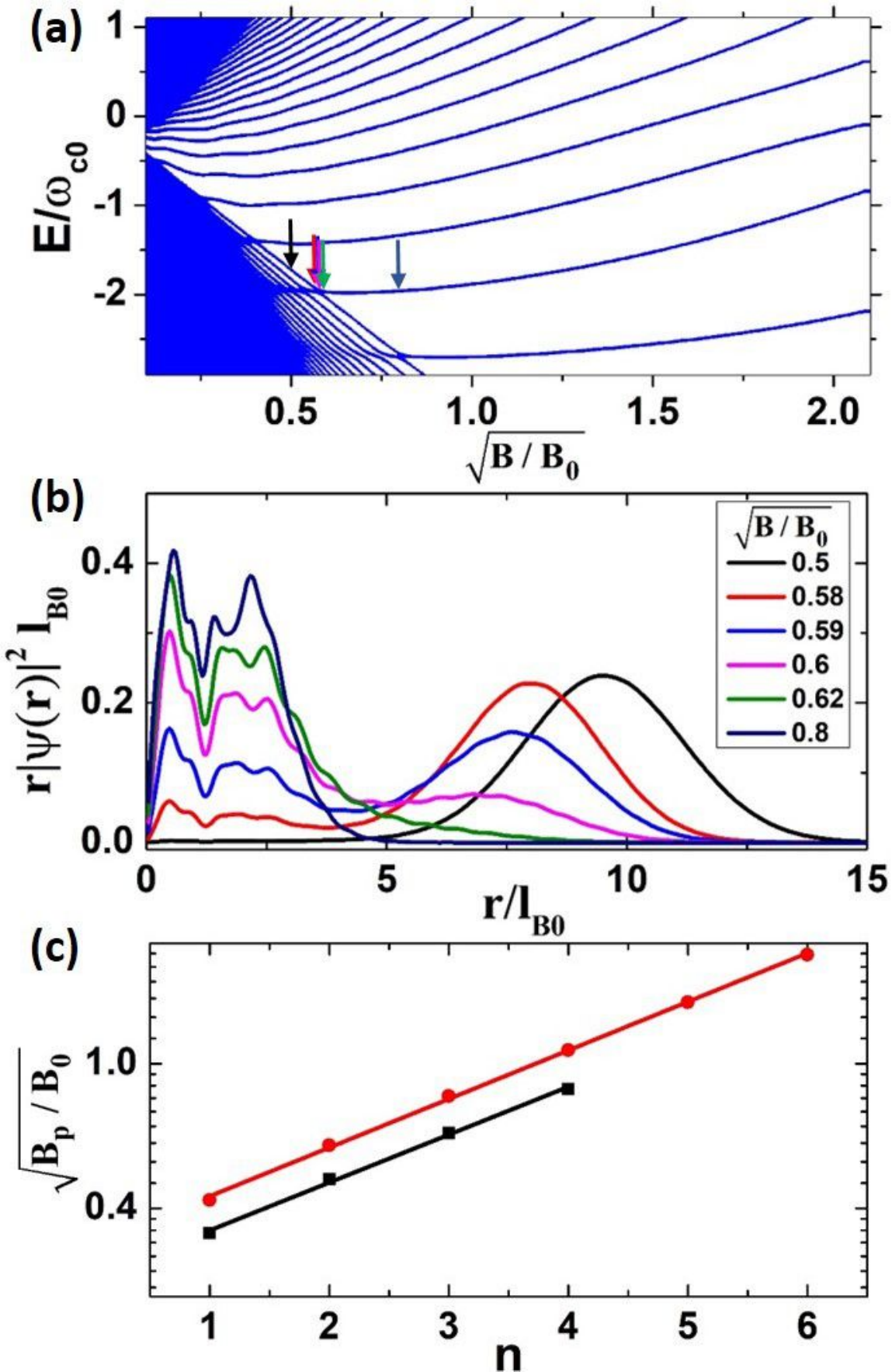}
    \caption{(Color Online). The spectrum of 2D massless Dirac electrons in an attractive Coulomb potential and varying magnetic field. A small width is added to make the spectrum visible. $Z\alpha=10$ is set. A magnetic field $B_0$ is chosen as a reference. $\omega_{c0}=\sqrt{2}\hbar v_F/l_{B0}$ is the cyclotron frequency of Dirac fermions. (a) The spectrum of the channel $m=0$. (b) The radial probability distributions of the energy branch (labeled by the colorful arrows in (a)) at various magnetic fields, which display a crossover from a Landau level to a bound state as the magnetic field increases. (c) Logarithmic plot of the magnetic positions (black dots), where the bound states start to form in the crossovers of different energy branches. A good linear fitting (black line) represent a geometric scaling law. Red dots denote the magnetic positions where the energy branches touch the Dirac point (i.e., E=0), as well as a linear fitting (red line), which have been well studied in Ref.~\cite{Sun}. The scaling factors for both are equal, which agrees well with the radial law in Efimov effect. }
   \label{Dirac2Dm0}
\end{figure}

\section{crossover in two-dimensional Dirac semimetals \label{sec2}}

In 2D, the massless Dirac equation for electrons in a Coulomb potential and a magnetic field can be simplified as
\begin{equation}
  v_F \vec{\sigma}\cdot (\mathbf{P}+e\mathbf{A}) \psi(\mathbf{r})=(E-V(r))\psi(\mathbf{r})
  \label{Dirac2D}
\end{equation}
where, $v_F$ is the Fermi velocity. $\vec{\sigma}$ are Pauli matrices and $\mathbf{P}$ is the momentum operator. $-e$ is the electric charge of the electron. $\mathbf{A}$ and $V(r)$ are respectively the magnetic vector potential and the Coulomb potential. We choose the symmetric gauge $\mathbf{A}=(-By/2,Bx/2,0)$. $V(r)=-Ze^2/(4\pi \varepsilon_0 r)$, with $Ze$ being the electric charge of the impurity and $\varepsilon_0$ being the vacuum permittivity. It can be written as $V(r)/(\hbar v_F)=-Z\alpha/r$ with the fine structure constant $\alpha=e^2/(4\pi\varepsilon_0\hbar v_F)$. $\psi(\mathbf{r})$ is the eigen-wavefunction with the eigen-value $E$.

In zero magnetic field, there exist a sequence of Efimov-like quasibound states with a scaling factor $\lambda = e^{\pi/\sqrt{(Z\alpha)^2-(m+1/2)^2}}$ when $Z\alpha>|m+1/2|$ ($m$ is an integer) \cite{AC4,AC6,AC12,DiracEfimov1,DiracEfimov2,DiracEfimov3,DiracEfimov4}. On the other hand, with no Coulomb potential, Landau levels form in magnetic field. Therefore, the combination of Coulomb potential and magnetic field leads to the interplay between quasibound states and Landau levels. 

We use the Landau level's wavefunctions as the basis to solve Eq.~\ref{Dirac2D} and obtain the energy spectrum as shown in Fig.~\ref{Dirac2Dm0}. Owing to the Coulomb potential, the Landau level does not satisfy $E_n \propto \sqrt{n}$ as the free Dirac fermions \cite{Dirac1,Sun}. Nevertheless, they roughly follow $E_n \propto \sqrt{B}$, as long as the size of the short-range boundary is negligible. When the Landau level meets the bound state, with varying magnetic field, there is an anti-crossing effect between two branches. The Landau level crossovers into the bound state for the upper branch, while the bound state crossovers into the Landau level for the lower branch. The magnetic positions for different crossovers obey a geometric scaling law, due to $E_n \propto \sqrt{B} \propto 1/l_B$ and the remaining scaling symmetry in the space-magnetic length domain. It is similar to the radial law of the Efimov effect, which is in the space-scattering length domain \cite{Efimov2}. Here, we employ the parameter $Z\alpha=10$ and for the channel $m=0$, our numerical value is $\lambda = 1.39$, which is in excellent agreement with the analytic one $\lambda=1.37$ in zero magnetic field.

\begin{widetext}

\begin{figure}[h] 
\includegraphics[width=17cm]{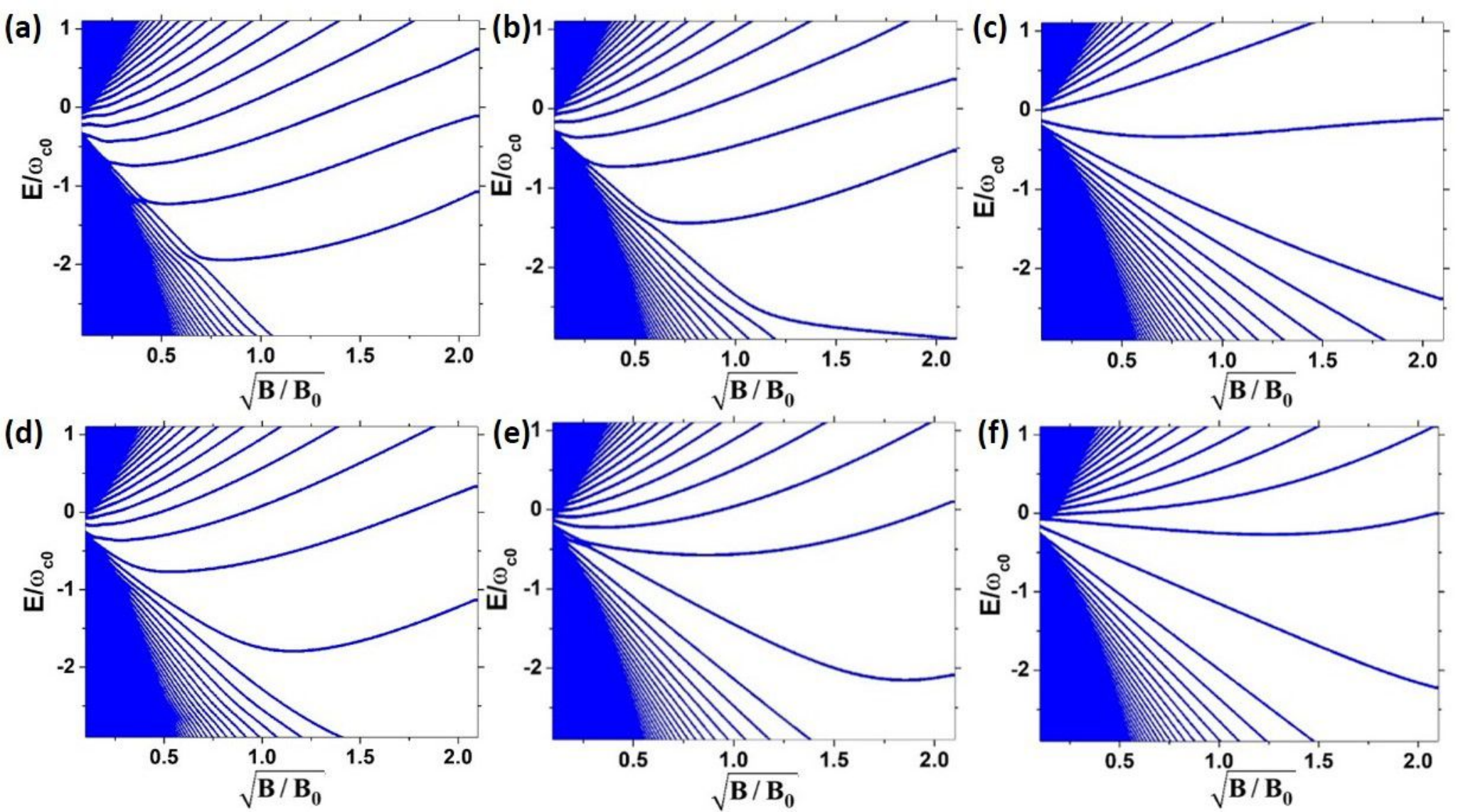}
\caption{The spectrum for different impurity charges $Z\alpha$=7(a), 5(b), 1(c) in the same channel $m=0$,  and different channels m=1 (d), 3(e), 5(f) with the same charge $Z\alpha=5$. Since the scaling factor is $e^{\pi/ \sqrt{(Z\alpha)^2-(m+1/2)^2}}$, both decreasing $Z\alpha$ and increasing $m$ can increase or even destroy it. Therefore, the crossover branches become sparser or even disappear.  } 
\label{fig3}
\end{figure}

\end{widetext}

Since the scaling factor $\lambda=e^{\pi/\sqrt{(Z\alpha)^2-(m+1/2)^2}}$, it becomes larger when we decrease $Z\alpha$ or increase $|m|$. Hence, the crossover branches become sparser, as displayed in Fig.~\ref{fig3}. When $Z\alpha<|m+1/2|$, they disappear as expected.

\section{crossover in two-dimensional three-component semimetals}

Three-component semimetals have been theoretically proposed and experimentally observed \cite{3S1,3S2,3S3,3S4,3S5,3S6,3S7,3S8,3S9}. Here, we study 2D massless three-component fermions with an attractive Coulomb potential and a magnetic field, as a generalization of 2D Dirac fermions in Sec.~\ref{sec2}. Similar to Eq.~\ref{Dirac2D}, the corresponding equation can be expressed as

\begin{equation}
  v_F \vec{S}\cdot (\mathbf{P}+e\mathbf{A}) \psi(\mathbf{r})=(E-V(r))\psi(\mathbf{r})
  \label{3S2D}
\end{equation}
Here, $\vec{S}$ are the generators of the rotation group SO(3) in the spin-1 representation, which replace the Pauli matrices in Eq.~\ref{Dirac2D}.

\begin{equation}
S_x=\frac{1}{\sqrt{2}}  \begin{pmatrix}
   0 & 1 & 0  \\
   1 & 0 & 1  \\
   0 & 1 & 0
 \end{pmatrix}, \ 
S_y=\frac{1}{\sqrt{2}}  \begin{pmatrix}
   0 & -i & 0  \\
   i & 0 & -i  \\
   0 & i & 0
 \end{pmatrix}
\end{equation}

\begin{figure}[b] 
    \includegraphics[width=8.0cm,height=13.0cm]{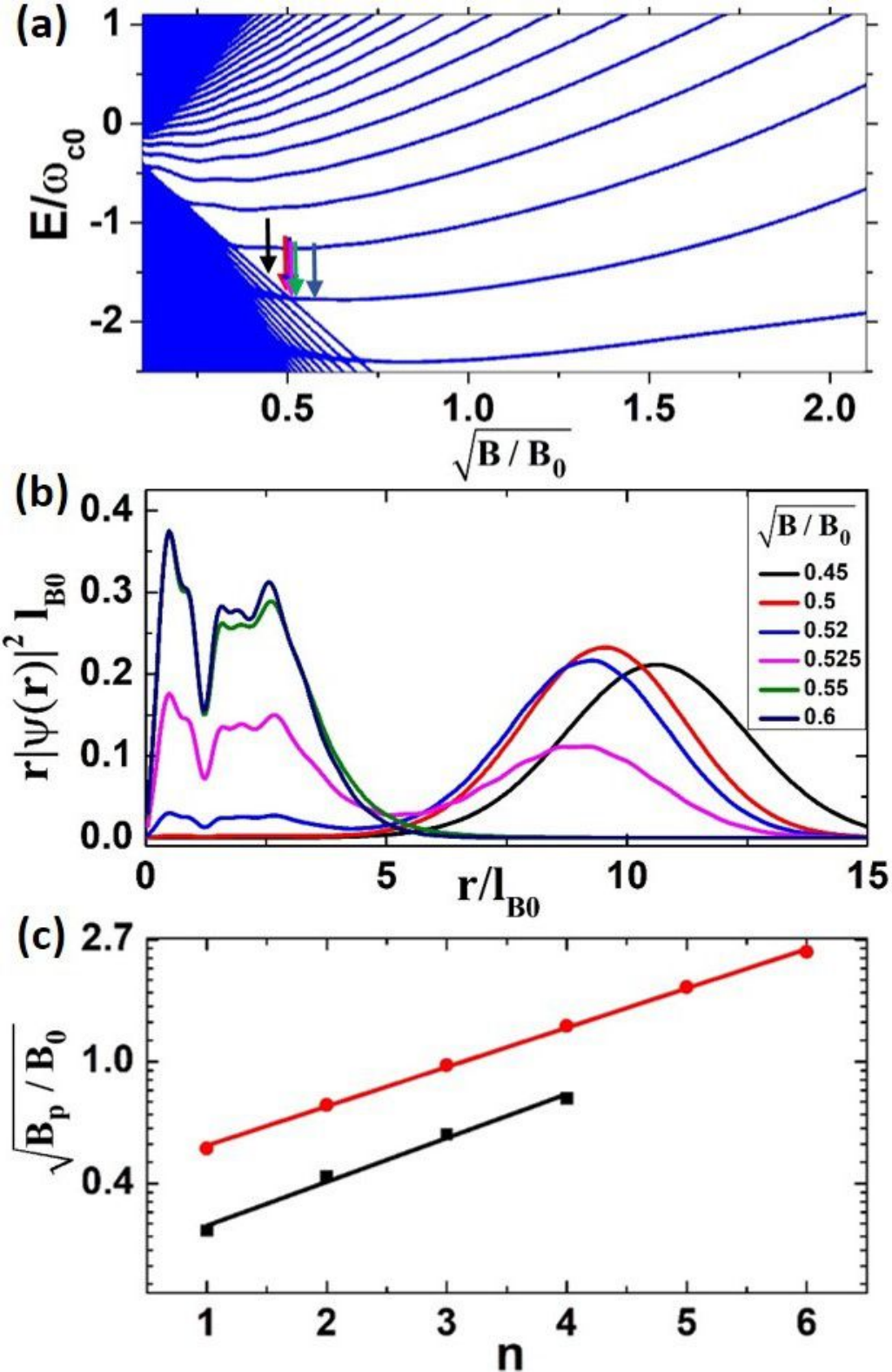}
    \caption{(Color Online). The spectrum of 2D massless three-component fermions in an attractive Coulomb potential and varying magnetic field. The same parameters are used as in Fig.~\ref{Dirac2Dm0}. (a) The spectrum of the channel $m=0$. (b) The radial probability distributions of the energy branch (labeled by the colorful arrows in (a)) at various magnetic fields, which display the same crossover from a Landau level to a bound state as the Dirac fermions. (c) Logarithmic plot of the magnetic positions (black dots) for the emerging bound states in the crossovers of different energy branches. Red dots denote the magnetic positions where the energy branches touch $E=0$. Both can be linearly fitted well and are in good agreement with the radial law. }
     \label{3S2Dm0}
\end{figure}

In zero magnetic field, if we assume $\psi(\mathbf{r})=(u_1(r)e^{i(m-1)\phi},u_2(r)e^{im\phi},u_3(r)e^{i(m+1)\phi})^T$, the radial equation can be written as

\begin{widetext}
\begin{equation}
\begin{pmatrix}
   -i\sqrt{2}(\frac{E}{\hbar v_F}+\frac{Z\alpha}{r}) & \frac{d}{dr}+\frac{m}{r} & 0  \\
   \frac{d}{dr}-\frac{m-1}{r} & -i\sqrt{2}(\frac{E}{\hbar v_F}+\frac{Z\alpha}{r}) & \frac{d}{dr}+\frac{m+1}{r}   \\
   0 &  \frac{d}{dr}-\frac{m}{r}  & -i\sqrt{2}(\frac{E}{\hbar v_F}+\frac{Z\alpha}{r})
 \end{pmatrix}
 \begin{pmatrix}
  u_1(r)  \\
  u_2(r)  \\
  u_3(r)
 \end{pmatrix}
 =0
\end{equation}
\end{widetext}
where, $m$ is an integer. It possesses a continuous scaling symmetry. When $Z\alpha>\sqrt{m^2+1/4}$, there exist Efimov-like quasibound states with a scaling factor $\lambda = e^{\pi/\sqrt{(Z\alpha)^2-m^2-1/4}}$,  similar to Dirac fermions. On the other hand, in a magnetic field with no Coulomb potential, Landau levels form with wavefunctions 
\begin{equation}
\frac{1}{\sqrt{2}}
 \begin{pmatrix}
  \mp i\sqrt{\frac{n+1}{2n+1}}\psi_{n+1,m-1}(\mathbf{r}) \\
  \psi_{n,m}(\mathbf{r})  \\
  \pm i\sqrt{\frac{n}{2n+1}}\psi_{n-1,m+1}(\mathbf{r})
 \end{pmatrix}
\end{equation}
for eigenvalues $E_n=\pm\sqrt{n+1/2}\hbar\omega_c$ ($n>0$ and $m\geq -n$. For $n=0$ Landau level, it is $1/\sqrt{2}(\mp i\psi_{1,m-1}, \psi_{0,m}, 0)^T$.)
\begin{align}
 \psi_{n,m}(\mathbf{r})= &\frac{1}{\sqrt{2\pi l_B^2}}\sqrt{\frac{n!}{(n+m)!}}e^{im\phi - r^2/(4 l_B^2)} \nonumber \\
  &\cdot(\frac{r^2}{2l_B^2})^{m/2} L_n^{(m)}(\frac{r^2}{2 l_B^2})
 \label{Landauwf}
\end{align}
Here, $L_n^{(m)}$ is the associated Laguerre polynomial and $\omega_c=\sqrt{2}\hbar v_F/l_B$ is the cyclotron frequency. We use these wavefunctions as the basis to solve Eq.~\ref{3S2D} and obtain the spectrum in magnetic field and Coulomb potential. The results are displayed in Fig.~\ref{3S2Dm0}. Here, we use $Z\alpha=10$, the same parameter as for Dirac fermions. So, for the channel $m=0$, their scaling factors are the same. Therefore, their spectra are very similar (see Fig.~\ref{Dirac2Dm0} and Fig.~\ref{3S2Dm0}). The slight difference results from the short-range boundary condition, even though they are also the same in our calculations. As expected, there are also Landau level to Efimov-like bound state crossovers. The numerical value of the scaling factor is $\lambda =1.38$, agreeing very well with the analytic one $\lambda=1.37$ in zero magnetic field. The comparison between three-component fermions and Dirac fermions indicates that the key requirement for the crossover is the existence of Efimov-like quasibound states in zero magnetic field, which is ubiquitous in 2D fermions with linear dispersion. Because the combination of linear dispersion and Coulomb potential make the corresponding equation exhibit a continuous scaling symmetry.

\section{Summary and Outlook}

In summary, 2D Dirac fermions can form Efimov-like quasibound states in an attractive Coulomb potential. When a magnetic field is applied, due to the gap between Landau levels, these quasibound states can transform into bound states, which satisfy a geometric scaling law in the space-magnetic length domain, an analogy of the radial law in the Efimov effect.  Landau level to Efimov-like bound state crossovers occur when the varying magnetic field makes them meet each other. This phenomenon can also appear in 2D massless three-component fermions and demonstrates the universality in 2D fermions with linear dispersion. Since Efimov-like quasibound states have already been observed in graphene \cite{DiracEfimov3}, our theoretical results can be explored experimentally in the near future.

Only one impurity was investigated in this paper. It will be interesting to consider a lattice of impurities, which interfere with each other and could lead to new phenomena. On the other hand, Hofstadter butterfly is expected in a periodic potential \cite{Hofstadter}. How Efimov-like bound (quasibound) states modify the structure of Hofstadter butterfly can be studied in the future.

\begin{acknowledgements} 
We thank Pengfei Zhang, Xin Chen, Zhigang Wu and Hong Yao for helpful discussions.
\end{acknowledgements}


\begin{thebibliography}{99}

\bibitem{Dirac1} A. H. Castro Neto, F. Guinea, N. M. R. Peres, K. S.
Novoselov, and A. K. Geim, Rev. Mod. Phys. {\bf 81}, 109 (2009).

\bibitem{Dirac2} M. Z. Hasan and C. L. Kane, Rev. Mod. Phys. {\bf 82}, 3045 (2010).

\bibitem{Dirac3} X. L. Qi and S. C. Zhang, Rev. Mod. Phys. {\bf 83}, 1057 (2011).

\bibitem{QHE} Y. Zhang, Y. W. Tan, H. L. Stormer and P. Kim, Nature {\bf 438}, 201 (2005).

\bibitem{AC1} Y. B. Zeldovich and V. S. Popov, Usp. Fiz. Nauk {\bf 105}, 403 (1971) [Sov. Phys. Usp. {\bf 14}, 673 (1972)].

\bibitem{AC2} W. Greiner, B. Muller, and J. Rafelski, \textit{Quantum Electrodynamics of Strong Fields} (Springer-Verlag, Berlin, 1985).

\bibitem{AC3} V. M. Pereira, J. Nilsson, and A. H. C. Neto, Phys. Rev. Lett. {\bf 99}, 166802 (2007).

\bibitem{AC4} A. V. Shytov, M. I. Katsnelson and L. S. Levitov, Phys. Rev. Lett. {\bf 99}, 246802 (2007).

\bibitem{AC5} A. V. Shytov, M. I. Katsnelson, and L. S. Levitov, Phys. Rev. Lett. {\bf 99}, 236801 (2007).

\bibitem{AC6} O.V. Gamayun, E.V. Gorbar, and V.P. Gusynin,
Phys. Rev. B {\bf 80}, 165429 (2009).

\bibitem{AC7} O.V. Gamayun, E.V. Gorbar, and V.P. Gusynin,
Phys. Rev. B {\bf 81}, 075429 (2010).

\bibitem{AC8} O.V. Gamayun, E.V. Gorbar, and V.P. Gusynin,
Phys. Rev. B {\bf 83}, 235104 (2011).

\bibitem{AC9} Y. Wang, D. Wong, A. V. Shytov, V. W. Brar, S. Choi, Q. Wu, H. -Z. Tsai, W. Regan, A. Zettl, R. K. Kawakami, S. G. Louie, L. S. Levitov, and M. F. Crommie, Science {\bf 340}, 734 (2013).

\bibitem{AC10} O.O.Sobol, E.V. Gorbar, and V.P. Gusynin, Phys. Rev. B {\bf 88}, 205116 (2013).

\bibitem{AC11} D. Moldovan, M. R. Masir and F. M. Peeters, 2D Mater. {\bf 5}, 015017 (2018).

\bibitem{AC12} E.V. Gorbar, and V.P. Gusynin, O.O.Sobol, Low
Temperature Physics {\bf 44}, 371 (2018).




\bibitem{DiracEfimov1} Y. Nishida, Phys. Rev. B {\bf 90}, 165414 (2014).

\bibitem{DiracEfimov2} Y. Nishida, Phys. Rev. B {\bf 94}, 085430 (2016).

\bibitem{DiracEfimov3} O. Ovdat, J. Mao, Y. Jiang, E. Y. Andrei, E. Akkermans, Nat. Commun. {\bf 8}, 507 (2017).

\bibitem{DiracEfimov4} P. F. Zhang and H. Zhai, Front. Phys. {\bf 13}, 137204 (2018).   


\bibitem{Efimov1} V. Efimov, {\it Yad. Fiz}. {\bf12}, 1080 (1970); Sov. J. Nucl. Phys. {\bf12}, 589 (1971). 

\bibitem{Efimov2} E. Braaten and H.-W. Hammer, Phys. Rep. {\bf 428}, 259
(2006).

\bibitem{Sun} M. Y. Sun, arXiv: 1805.02074.



\bibitem{3S1} T. T. Heikkilä and G. E. Volovik, New J. Phys. {\bf 17}, 093019 (2015).

\bibitem{3S2} B. Bradlyn, J. Cano, Z. Wang, M. G. Vergniory, C. Felser, R. J. Cava, and B. A. Bernevig, Science {\bf 353}, aaf5037 (2016). 

\bibitem{3S3} G. W. Winkler, Q. Wu, M. Troyer, P. Krogstrup, and A. A. Soluyanov, Phys. Rev. Lett. {\bf 117}, 076403 (2016).

\bibitem{3S4} T. Hyart and T. T. Heikkilä, Phys. Rev. B {\bf 93},
235147 (2016).

\bibitem{3S5} H. Weng, C. Fang, Z. Fang, and X. Dai, Phys. Rev. B {\bf 93}, 241202 (2016).

\bibitem{3S6} Z. Zhu, G. W. Winkler, Q. S. Wu, J. Li, and A. A. Soluyanov, Phys. Rev. X {\bf 6}, 031003 (2016).

\bibitem{3S7} H. Weng, C. Fang, Z. Fang, and X. Dai, Phys. Rev. B {\bf 94}, 165201 (2016).

\bibitem{3S8} G. Chang, S.-Y. Xu, S.-M. Huang, D. S. Sanchez, C.-H. Hsu, G. Bian, Z.-M. Yu, I. Belopolski, N. Alidoust, H. Zheng, T.-R. Chang, H.-T. Jeng, S. A. Yang, T. Neupert, H. Lin, and M. Z. Hasan, 	Sci. Rep. {\bf 7}, 1688 (2017).  

\bibitem{3S9} B. Q. Lv, Z.-L. Feng, Q.-N. Xu, J.-Z. Ma, L.-Y. Kong, P. Richard, Y.-B. Huang, V. N. Strocov, C. Fang, H.-M. Weng, Y.-G. Shi, T. Qian, H. Ding, Nature (London) {\bf 546}, 627 (2017).


\bibitem{Hofstadter} D. Hofstadter, Phys. Rev. B {\bf 14}, 2239 (1976).



%
%
%
%
%
%
%
%
%
%
%
%
%
%
%
%
%
%
%
%
%
%
%



\end{thebibliography}
\end{document}